\documentclass[prl,twocolumn,preprintnumber,showpacs]{revtex4}
\usepackage{graphicx}
\usepackage{amsmath}
\usepackage{amssymb}
\usepackage{epsf}
\usepackage{natbib}

\begin{document}

\title{Temporally heterogeneous dynamics in granular flows}

\author{Leonardo E.~Silbert}\email{lsilbert@uchicago.edu}

\affiliation{James Franck Institute, University of Chicago, Chicago,
  Illinois 60637} \affiliation{Department of Chemistry and
  Biochemistry, University of California Los Angeles, Los Angeles,
  California 90095} 

\date{\today}

\begin{abstract}
  
  Granular simulations are used to probe the particle scale dynamics
  at short, intermediate, and long time scales for gravity driven,
  dense granular flows down an inclined plane. On approach to the
  angle of repose, where motion ceases, the dynamics become
  intermittent over intermediate times, with strong temporal
  correlations between particle motions -- \textit{temporally
    heterogeneous dynamics}. This intermittency is characterised
  through large scale structural events whereby the contact network
  periodically spans the system. A characteristic time scale
  associated with these processes increases as the stopped state is
  approached. These features are discussed in the context of the
  dynamics of supercooled liquids near the glass transition.

\end{abstract}


\maketitle 

Granular materials persist at the forefront of contemporary research
due to the extreme richness in their dynamics, either under shear
\cite{losert2}, flowing out of a hopper \cite{bazant1}, or driven by
gravity \cite{losert1}. Because thermodynamic temperature plays little
role in determining these features, granular materials are widely
recognized as a macroscopic analogue of athermal systems far from
equilibrium \cite{kadanoff1}. The notion of \textit{jamming}
\cite{liu1} provides a unifying framework to describe the behaviour of
a wide range of systems, from the molecular level (supercooled liquids
and glasses), microscale (colloidal glasses), through to the
macroscale (granular matter), near their jamming transition. The
motivation for this work comes from the view that granular flows {\em
  can} be used as a macroscopically accessible analogue through which
we can gain a better understanding of the critical slowing down of the
dynamics in systems near their point of jamming. One of the features
associated with the approach of the glass transition is the concept of
spatially heterogeneous dynamics \cite{ediger2}. This work presents
features of a driven granular material flowing down an inclined plane
-- chute flows -- that exhibit large scale, cooperative events on
approach to the angle of repose where flow stops -- its jamming point.

Gravity driven flows continue to be studied extensively due to their
ubiquity throughout nature -- avalanches, debris flows, and sand dunes
-- and also because of their importance in materials handling
throughout industry. More recently, there has been a focus of
attention towards understanding the properties of chute flows because
of the ability of obtaining well-defined and reproducible steady state
flow regimes. Despite the fact that recent experiments have been able
to determine, in detail, macroscopic information of granular flows
\cite{pouliquen1,azanza1}, probing the internal dynamics remains
problematic due to the opaqueness of the particles. Still, such flows
offer a relatively clean system in which to develop theoretical models
for constitutive relations as well as describing the flow at the level
of the grain size \cite{mills4}. This is where the role of simulation
has proved extremely useful.

\begin{figure}[h]
\begin{center}
  \includegraphics[width=7cm,height=1.5cm]{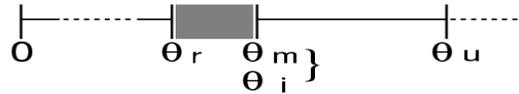}
\caption{Phase behaviour of granular chute flow (schematic). $\theta_{r}$
  is the angle of repose at which flow ceases. $\theta_{m}$ is the
  maximum angle of stability at which flow is initiated from rest.
  $\theta_{u}$ is the maximum angle for which stable flow is observed.
  The angle $\theta_{i}$ at which intermittent dynamics emerges, the
  shaded region, coincides with $\theta_{i}\approx\theta_{m}$. For the
  simulations reported here, $\theta_{r}=19.30^{\circ}\pm
  0.01^{\circ}$, $\theta_{m} = 20.3^{\circ}\pm 0.2^{\circ}$, and
  $\theta_{u}\approx 26^{\circ}$.}
\label{fig1}
\end{center}
\end{figure}

Granular simulations of chute flows have captured features of the
rheology with remarkable accuracy \cite{leo7,leo12}. Three principal
flow regimes are observed for chutes with rough bases (see
Fig.~\ref{fig1}). If $\theta$ is the inclination angle of the chute:
i) no flow occurs below the angle of repose $\theta_{r}$, ii) steady
state flow exists in the region $\theta_{r}<\theta<\theta_{u}$, and
iii) unstable flow occurs for $\theta>\theta_{u}$. For all
$\theta<\theta_{u}$, the density is constant throughout the bulk, away
from the free top surface and lower boundary, with smoothly varying
depth profiles of the flow velocity and the principle and shear
stresses \cite{leo7}. The location of $\theta_{r}$ also depends on the
height of the flowing pile $h$ \cite{pouliquen1,leo12}. Pouliquen
introduced the quantity $h_{stop}(\theta)$ that encodes all the
(undetermined) information about the roughness and effective friction
of the base to relate the dependency of $\theta_{r}$ on $h$. In the
language of glasses and colloids, $h_{stop}(\theta)$ represents the
glass transition or yield stress line, respectively.

This work characterises the large-scale fluctuations and emerging
spatial and temporally heterogeneous dynamics for a flowing state in
the vicinity of $\theta_{r}$. This is in contrast to discrete
avalanches in heap flows \cite{durian1}. $\theta_{i}$ is the angle,
for $\theta_{r} < \theta \leq \theta_{i}$, such that, over time scales
intermediate between ballistic propagation, at short times, and
diffusive motion, at long times, the particle dynamics become strongly
intermittent. This intermittency is a consequence of large spatial and
temporal correlations in particle motions, through the relaxation and
reformation of mechanically stable clusters that periodically span the
system. A characteristic time scale $\tau_{c}$, associated with these
correlations is found to increase as $(\theta - \theta_{r})
\rightarrow 0$.

The simulations were performed using the discrete element technique
with interactions appropriate for granular materials \cite{leo7}.
Monodisperse (soft) spheres of diameter $d$ and mass $m$ flow down a
roughened, inclined plane due to gravity $g$. Initially, $\theta$ is
set to $\theta_{u}$ to remove any preparation history, then
incrementally reduced in steps of $0.01^{\circ} \leq \Delta\theta \leq
0.5^{\circ}$. Particles interact only on contact (cohesionless), along
directions normal and tangential (via static friction) to the vector
joining their centres. Contacts are defined as two overlapping
neighbours. All runs were performed with a particle friction
coefficient $\mu =0.5$ and coefficient of restitution $\epsilon =
0.88$. Most of the results presented are for $N=8000$ particles
flowing down a chute that is $L_{x}=20d$ long, $L_{y}=10d$ wide (the
$xy$-plane employs periodic boundary conditions), and an average flow
height $h\approx 40d$ (with a free top surface). See Fig.~\ref{fig5}.
Massively, large-scale, parallel simulations were used for $N=160000$,
with $(L_{x},L_{y},h) = (100,40,40)d$, to demonstrate that the
observed features are not a system size artifact. Length scales are
non-dimensionalised by the particle diameter $d$ and time in units of
$\tau=\sqrt{d/g}$.

The intermittent regime was first identified through measurements of
the kinetic energy as shown in Fig.~\ref{fig1}. In the
continuous-flow, steady state regime the energy is approximately
constant with small fluctuations about the mean value.  As the
inclination angle is incrementally decreased towards $\theta_{r}$,
changes in the time profile of the energy first become apparent for
$\theta_{i}\approx 20.2^{\circ}$.
\begin{figure}[h]
\begin{center}
  \includegraphics[width=8cm]{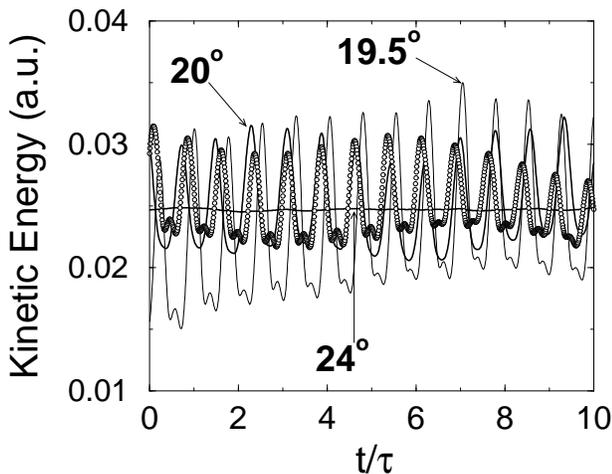}
\caption{Average kinetic energy per particle over an intermediate time window
  for $N=8000$ and different inclination angles as indicated. The
  symbols are for $N=160000$ at $\theta=19.5^{\circ}$.}
\label{fig2}
\end{center}
\end{figure}

\begin{figure}[h]
\begin{center} 
  \includegraphics[width=8cm]{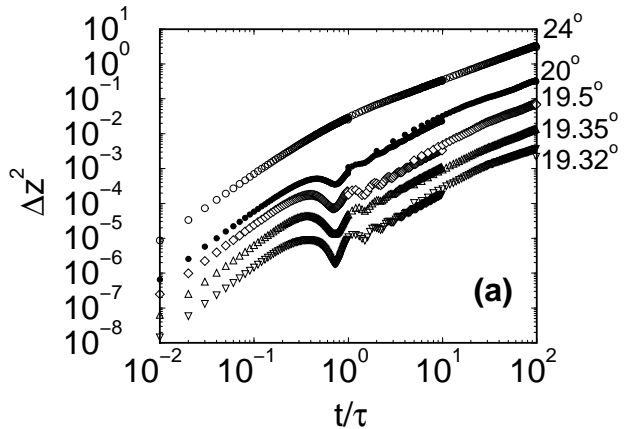}
\caption{The mean squared displacement in the direction normal to the
  plane $\Delta z^{2}$, with decreasing angle from top to bottom. All
  curves have been shifted for clarity.}
\label{fig3}
\end{center}
\end{figure}

\begin{figure}[h]
\begin{center}
  \includegraphics[width=8cm]{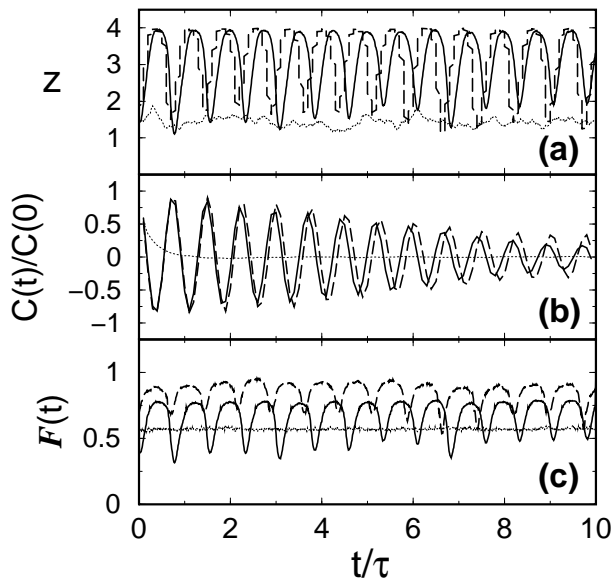}
\caption{Intermediate time window associated with the oscillations in the MSD,
  for $19.35^{\circ}$ (dashed line), $19.5^{\circ}$ (thick), and
  $24^{\circ}$ (dotted): (a) coordination number $z$, (b) velocity
  autocorrelation function $C(t)$, and (c) force-force time
  correlation function $\mathcal{F}(t)$.}
\label{fig4}
\end{center}
\end{figure}

\begin{figure}[h]
\begin{center}
  \includegraphics[width=2.5cm,height=4cm]{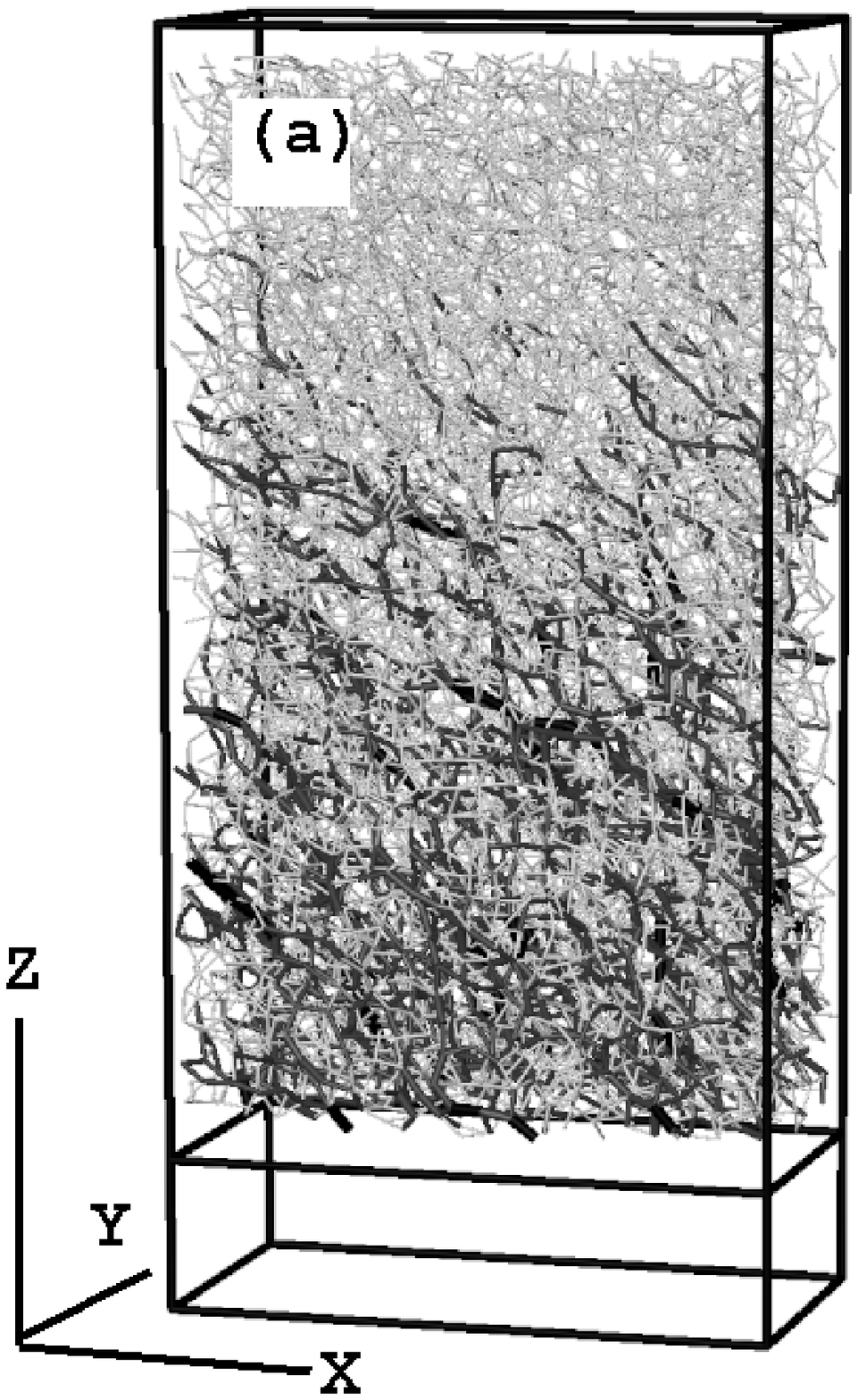}
  \includegraphics[width=2.5cm,height=4cm]{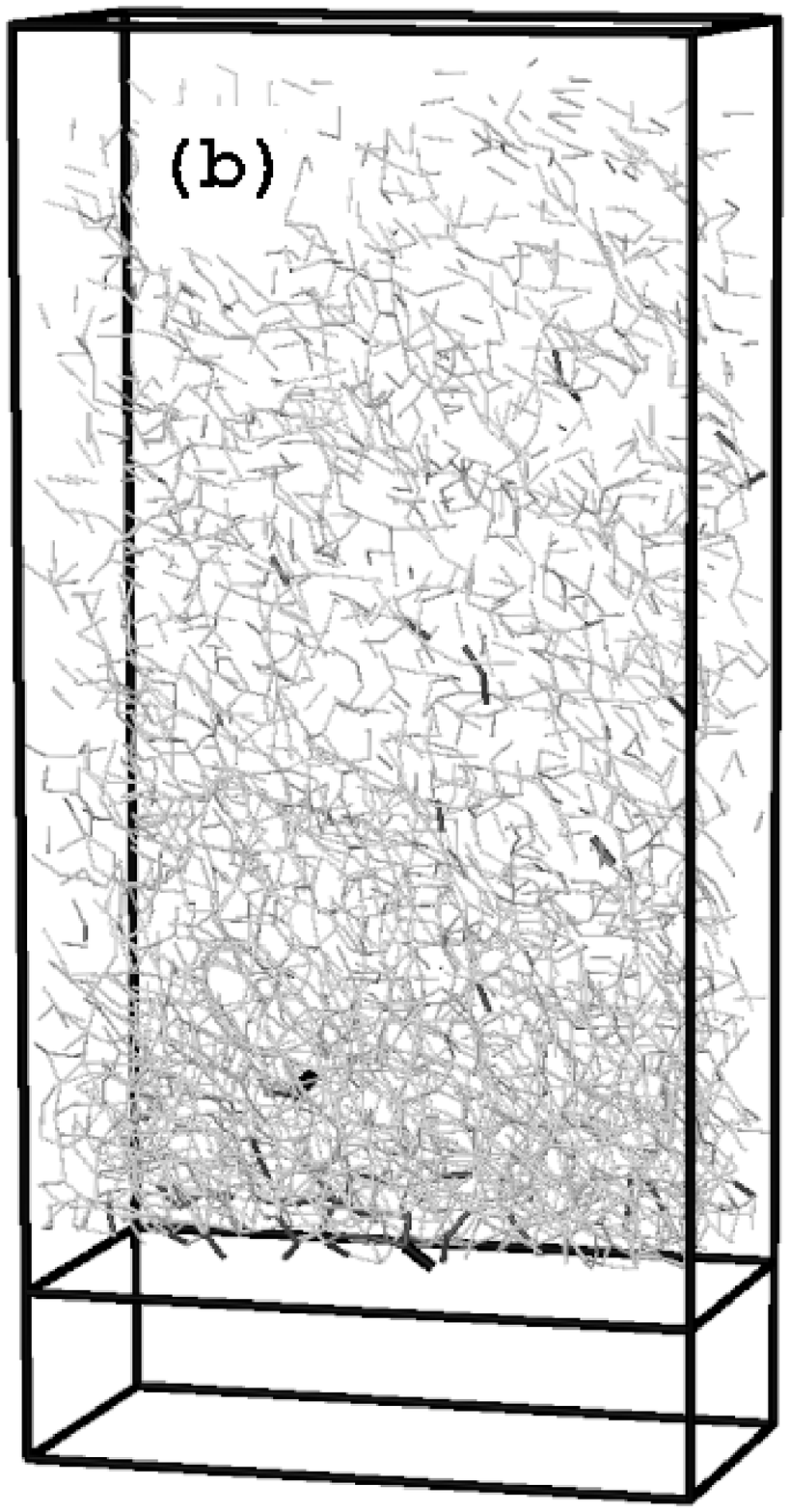}
  \includegraphics[width=2.5cm,height=4cm]{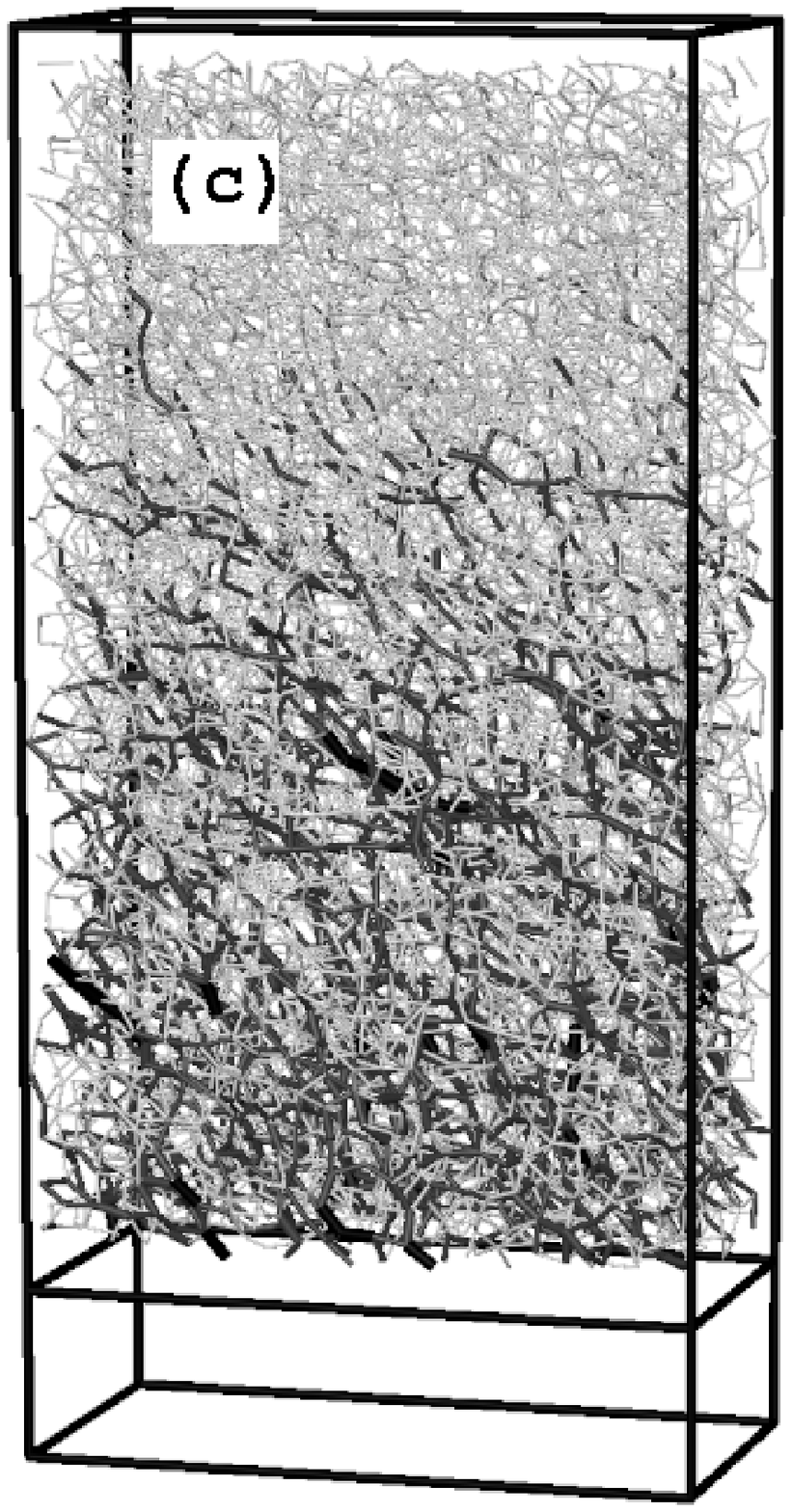}
\caption{The contact force network at $\theta=20^{\circ}$ for three successive
  times. (a) Snapshot corresponds to a peak in the coordination
  number, (b) a trough, and (c) the next peak. The shade and thickness
  of the lines represent the magnitudes of the normal forces between
  two particles in contact. The inclination has been removed in the
  figures and the frame is a guide to the eye. The simulation axes are
  indicated.}
\label{fig5}
\end{center}
\end{figure}

Because of the convective nature of the flow, the mean-squared
displacement (MSD), shown in Fig.~\ref{fig3}, is measured normal to
the inclined plane in the $z$ direction: $\Delta z^{2} \equiv
<(z(t)-z(0))^{2}>$. During continuous flow, there is a clear crossover
between the ballistic regime, for $t/\tau < 10^{-1}$, and diffusive
regime, $t/\tau > 10$. Closer to the jamming point, the MSD oscillates
over times intermediate between them.
The appearance of novel dynamics in the vicinity of $\theta_{r}$ is a
consequence of strong correlations emerging in time-time quantities
such as the velocity autocorrelation function, $C(t) \equiv
<v_{z}(t)v_{z}(0)>$, where $v_{z}$ is the velocity component in the
direction normal to the plane. See Fig.~\ref{fig4}(b). These
fluctuating quantities are a signature of temporally heterogeneous
dynamics. To investigate this further, the average coordination number
$z$, Fig.~\ref{fig4}(a), is shown over the intermediate time window,
demonstrating that the structural processes associated with these
correlations involve large-scale cooperative events. During the {\em
  flowing} phase of the intermittent regime, $z$ remains lower than
the mechanically stable limit for frictional spheres $z_{c} = 4$
\cite{edwards2}. In the {\em static} phase, the systems attains a
value $z\approx z_{c}$, thus almost coming to rest.  (There is always
a residual creeping motion as flow does not completely stop until
$\theta_{r}$.)

To facilitate such processes, the contact force network undergoes
temporal transitions between static and flowing phases. This evolution
is captured in the (distinct) particle-particle contact force-force
time correlation function $\mathcal{F}(t)$, as shown in
Fig.\ref{fig4}(c). The physical picture of the intermittent regime is
thus: over some characteristic time scale the system oscillates
between almost total mechanical stability and fluid-like flow. This
involves time-dependent, system-spanning, structural relaxation events
through the break-up and reformation of particle contacts. This
process is best depicted in the simulation snapshots of
Fig.~\ref{fig5}.

\begin{figure}[h]
\begin{center}
  \includegraphics[width=8cm]{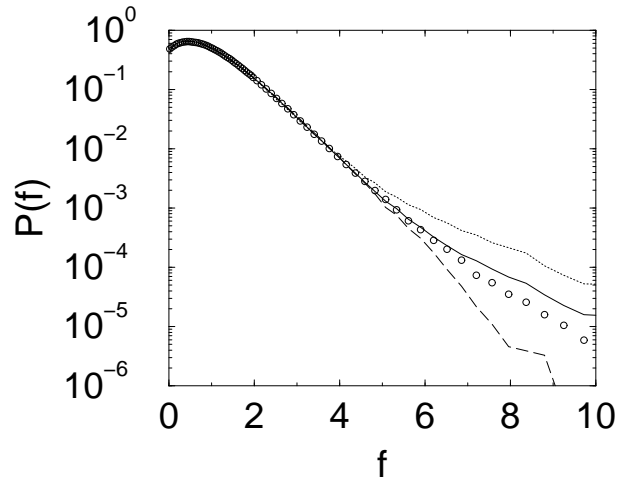}
  \caption{Probability distribution function $P(f)$ of the normal contact
    forces (normalised so $\bar{f}=1$) for $\theta=20^{\circ}$; total
    (solid line), the high-$z$ phase (dashed), and the low-$z$ phase
    (dotted). In the high-$z$ phase, the contact forces are binned
    into a histogram if $z>3.5$ for that configuration, and similarly
    if $z<2.0$ in the low-$z$ curve. The full $P(f)$ for $\theta
    =22^{\circ}$ (symbols) is shown for comparison.}
\label{fig6}
\end{center}
\end{figure}

The intermittent regime can therefore be considered as temporally
biphasic. During one phase, $z \approx z_{c}$, so the system is close
to mechanical stability, while the flowing phase has a considerably
lower coordination. A signature of this is seen in the probability
distribution function of the normal contact forces $P(f)$.  The
partial distributions shown in Fig.~\ref{fig6} are for the high-$z$
phase (defined as configurations with $z>3.5$) which resembles that of
a static packing -- exponential decay at high forces -- and separately
the low-$z$ phase ($z<2$), with a high-$f$ tail that decays more
slowly. This suggests that it may be possible to distinguish between
different types of flows depending on the time window of observation.

The intermittent regime is peculiar due to the complex interplay
between several different characteristic length and time scales.
Identifying the structural entities of characteristic size $\ell$,
remains a matter of debate \cite{mills4,leo14}. However, the general
picture is as follows: one expects $\ell$ to increase with decreasing
$\theta$, or correspondingly, the shear rate $\dot{\gamma}$. Then,
$\theta_{r}$ is the angle where $\ell \approx h$, resulting in the
transition from a dynamic state to a static one. The development of an
intermittent regime suggests another competing time scale, whereby
$\ell \approx h$ can occur, but other mechanisms drive the system away
from stability before the system permanently jams. This seems to
coincide with the regime where the scaling of the flow velocity
differs from that of the continuous flow regime \cite{leo12}.
Presently, it is not clear how to approach this problem without a more
sophisticated particle scale analysis.

\begin{figure}[h]
\begin{center}
  \includegraphics[width=8cm]{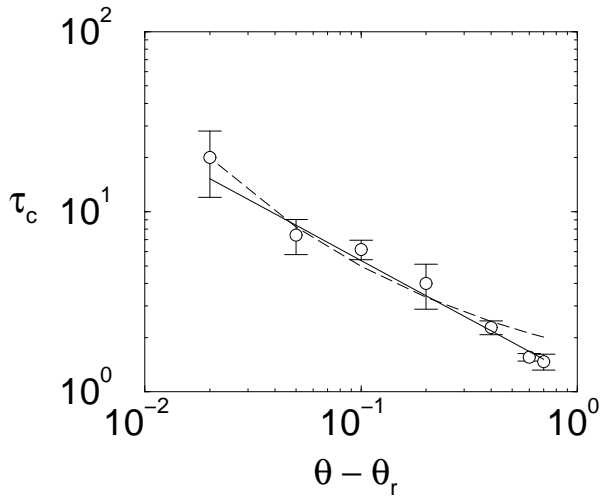}
\caption{Characteristic time scale $\tau_{c}$, with error bars, associated with
  the intermittent structural events at intermediate times as a
  function of distance from the jamming point $(\theta-\theta_{r})$.
  The solid line is a power law fit: $\tau \sim (\theta -
  \theta_{r})^{-\gamma}$, with $\gamma \approx 0.65$; whereas the
  dashed line is of the modified Vogel-Fulcher form: $\tau_{c} \sim
  \exp \{A(\theta-\theta_{r})^{-p}\}$, with $A\approx 0.66$ and $p
  \approx 0.39$.}
\label{fig7}
\end{center}
\end{figure}

Progress can still be made by identifying the characteristic time
scales of the intermittent regime. One is the periodicity in the
kinetic energy, correlation functions, and $z$. This time scale is
rather insensitive to $\theta$ in the intermittent phase. However,
$C(t)$ has an envelope of decay that corresponds to the extent of the
intermittency. A time scale $\tau_{c}$, that characterises the initial
decay in $C(t)$ is shown in Fig~\ref{fig7}. Interestingly, a power-law
divergence: $\tau_{c} \sim (\phi - \phi_{r})^{-\gamma}$, with $\gamma
\approx 0.65$, captures the behaviour quite well. As a comparison, a
modified Vogel-Fulcher form is also shown in Fig.~\ref{fig7}, which is
satisfactory close to the jamming point, but deviates further away.
The available range for analysis is somewhat restricted through this
particular definition of $\tau_{c}$, so it remains unclear how this
time scale might correspond to the more familiar definitions of
relaxation times in supercooled fluids \cite{angell1}. A more
appropriate definition of $\tau_{c}$ is being investigated.

In conclusion, gravity-driven, dense, granular flows down an inclined
plane exhibit intermittent dynamics over intermediate time scales in
the vicinity of the angle of repose. The intermittent regime signals
the onset of temporal heterogeneous dynamics, whereby the short time
motion is ballistic, at long times diffusive, but at intermediate
times a combination of creeping flow and trap-like dynamics exists.
The system becomes trapped in temporally metastable states
\cite{bouchaud1} through large scale cooperative events. Correlations
between these events decay, but do so more slowly as the jamming point
is approached. The behaviour of the characteristic time scale
$\tau_{c}$, shares similarities with the properties of supercooled
liquids in the vicinity of the glass transition. Such dynamical
similarities were recently probed experimentally in vibrated bead
packs \cite{danna2}. Therefore, in a typically heuristic fashion, an
analogy with glassy dynamics theory can be made. $\theta_{r}$ marks
the point where all dynamics cease and in this sense plays the role of
the glass transition temperature $T_{g}$. Indeed, earlier work
\cite{leo10} showed that the flowing-static transition resembles the
way a model liquid approaches the glassy phase \cite{ohern1}.
However, the way the dynamics change at $\theta_{i}$, is similar is
spirit to the way the dynamics in a supercooled fluid changes at the
mode coupling temperature $T_{c}$ \cite{gotze2}. With this view in
mind, one can associate $\theta_{i}$ with $T_{c}$. These results
therefore hint towards a unifying picture of dynamical heterogeneities
in dense, amorphous systems.  Because the time scales associated with
the onset of dynamical heterogeneity and cooperativity are
experimentally accessible, granular flows offer a convenient system to
address questions often associated with glasses at the molecular and
colloidal level. This emerging picture thus begs the question, can
existing theories that are currently being applied to traditional
glasses and colloids be extended to granular systems?

I thank Bulbul Chakraborty for insightful conversations and Gary S.
Grest for critiquing the manuscript. Grant support from Grant
Nos.~NSF-DMR-0087349, DE-FG02-03ER46087, NSF-DMR-0089081, and
DE-FG02-03ER46088 is gratefully acknowledged.

\end{document}